\documentclass[journal]{IEEEtran}
\usepackage[left=1.91cm, right=1.91cm, top=2.51cm]{geometry}
\usepackage{balance}
\usepackage{pbox}
\usepackage{enumerate}
\usepackage{cite}
\usepackage{graphicx}
\usepackage{graphicx, subfigure}
\usepackage{alltt}
\usepackage{amsmath}
\usepackage{amsopn}
\usepackage{amssymb}

\newtheorem{lemma}{Lemma}
\newtheorem{proposition}{Proposition}

\newtheorem{definition}{Definition}
\newtheorem{claim}{Claim}

\DeclareMathOperator{\GF}{GF}
\DeclareMathOperator{\prob}{Pr}

\title{Flexible Fractional Repetition Codes for Distributed Storage Networks}
\newcommand{\nkdset}{\mathcal{S}}
\author{Imad~Ahmad,~\IEEEmembership{Member,~IEEE,}
        and~Chih-Chun~Wang,~\IEEEmembership{Senior Member,~IEEE}
\thanks{This work was supported in part by NSF grants ECCS-1407603, CCF-1422997, and CCF-1618475.}
\thanks{I. Ahmad is currently with AT\&T Labs \{imadfahmad@gmail.com\} and C.-C. Wang is with the School of Electrical and Computer Engineering, Purdue University, West Lafayette \{chihw@purdue.edu\}.}
}

\begin{document}
\maketitle
\thispagestyle{plain}
\pagestyle{plain}

\begin{abstract}
Consider the following fundamental question of distributed storage networks: Given any arbitrary $(n,k,d)$ values, whether there exists an intelligent helper selection scheme (assuming unlimited memory and computing power) that can strictly improve the storage-bandwidth (S-B) tradeoff. Ahmad {\em et al.} 18' answered this question by proving that for a subset of $(n,k,d)$ values, no helper selection scheme can ever improve the S-B tradeoff, and for the $(n,k,d)$ not in that subset, a new scheme called family helper selection (FHS) can strictly improve the S-B tradeoff over a {\em blind} helper selection scheme. Nonetheless, the analysis of FHS is done by a min-cut analysis with no actual code construction. 
\par This work fills this gap between pure min-cut analysis and actual code construction by pairing FHS with a new, generalized version of the existing fractional repetition (FR) codes.  Specifically, existing FR codes are exact-repair codes that admit the highly-desirable
{\em repair-by-transfer} property, but its unique construction limits the application to a restricted set of $(n,k,d)$ values. In contrast, our new construction, termed \emph{flexible fractional repetition codes}, can be applied to arbitrary $(n,k,d)$ while retaining most of the practical benefits of FR codes, i.e., admitting small repair bandwidth, being exact-repair, and being {\em almost} repairable-by-transfer.
\end{abstract}

\section{Introduction} \label{sec:intro}
\par \IEEEPARstart{C}{onsider} the distributed storage network (DSN) formulated in \cite{dimakis2010network}. In \cite{ahmad2018can}, Ahmad \emph{et al.} identified a set of $(n,k,d)$ parameters, denoted by $\nkdset$, for which an optimally designed helper selection scheme can achieve strictly better storage-bandwidth (S-B) tradeoff than the {\em blind} helper selection (BHS) scheme originally proposed in \cite{dimakis2010network}. The results in \cite{ahmad2018can} also proved the corresponding converse: That is, for any $(n,k,d)\notin\nkdset$, no helper selection scheme can do better than BHS, i.e., BHS is already optimal.  The results in \cite{ahmad2018can} thus answer a fundamental question: Under what $(n,k,d)$ values can an intelligent helper selection scheme improve the performance of a DSN?

\par This work is motivated by an important code design problem that was omitted in the achievability results of \cite{ahmad2018can}. Specifically, \cite{ahmad2018can} devised a helper selection scheme termed \emph{the family helper selection (FHS)} scheme, characterized its S-B tradeoff for those $(n,k,d)\in\nkdset$, and showed that the corresponding S-B curve is strictly better than that of BHS. Specifically, the S-B tradeoff curve of FHS was derived using a graph-based analysis that quantifies the minimum possible min-cut value of FHS without any actual code construction. In a way similar to \cite{dimakis2010network}, the approach in \cite{ahmad2018can} assumed that there exist network codes that can achieve the min-cut-based S-B tradeoff without any discussion whether/how such a code can be constructed. Unfortunately, such a widely used assumption (see \cite{dimakis2010network}) represents a missing link in a truly rigorous DSN analysis. For example, whether there exists an S-B-curve achieving code depends heavily on the underlying finite field $\GF(q)$ and on the sub-packetization levels of the construction. None of these important code attributes is carefully analyzed in the min-cut analysis of \cite{dimakis2010network,ahmad2018can}. Whether there exists a large but fixed $\GF(q)$ that attains the S-B tradeoff curve is a non-trivial problem in the DSN literature since the results in \cite{ahlswede2000network,li2003linear} only guarantee the existence of $\GF(q)$ broadcast network codes for any {\em network/graph of bounded size}, but the size of the information flow graph (IFG) of a distributed storage code (see \cite{dimakis2010network} or \cite{ahmad2018can} for the description of these graphs) is unbounded. Therefore, the conclusion in \cite{ahlswede2000network} and \cite{li2003linear} does not imply the code existence. 

\par The original min-cut analysis of regenerating codes (RCs) in \cite{dimakis2010network} is complemented by a follow-up work \cite{wu2010existence}, which proved that there exists a fixed alphabet $\GF(q)$ and a corresponding network code that achieves the graph-based S-B tradeoff characterized in \cite{dimakis2010network}. Specifically, it outlined a detailed code construction that achieves all the points on the S-B tradeoff curve of RCs \cite{dimakis2010network} and the codes provided in \cite{wu2010existence} fall under the category of \emph{functional-repair} codes. Subsequent development in this direction (constructing codes that attain the graph-based S-B tradeoff) has been focused on \emph{exact-repair} codes \cite{shah2012interference,rashmi2011optimal,cadambe2013asymptotic,wu2011construction,shah2012distributed,wu2009reducing,el2010fractional}, for which during repair, the newcomer has to restore the data that was originally stored on the failed node. E.g., an exact-repair code construction, called {\em product-matrix} construction, that achieves the minimum-bandwidth-regenerating (MBR) point of RCs was proposed in \cite{rashmi2011optimal}. Another type of exact-repair codes called \emph{fractional repetition (FR) codes} was proposed in \cite{el2010fractional}, which, additionally, admits an important practical property called {\em repair-by-transfer}. 

\par Each of the product-matrix codes and the FR codes has its own distinct advantage. For example, the product-matrix codes \cite{rashmi2011optimal} can naturally \emph{handle multiple failures} since it guarantees that the newcomer can repair from \emph{any} set of $d$ surviving nodes. In contrast, the FR codes rely on the concept of {\em repetition} and thus were originally designed for the single-failure scenario. There are new generalizations of FR codes for multiple failures \cite{el2010fractional,koo2011scalable,olmez2016fractional,olmez2013replication}, but they are at the cost of decreasing the performance (not necessarily achieving the MBR point of BHS anymore) and further restricting the applicable $(n,k,d)$ values. On the other hand, the repair-by-transfer property of FR codes allows each helper to send a subset of the packets they store (without mixing them) to the newcomer, which significantly reduces the encoding complexity and, perhaps more importantly, the disk/memory I/O.

\par
\begin{figure}[h!]
\centering
\includegraphics[width=0.47\textwidth]{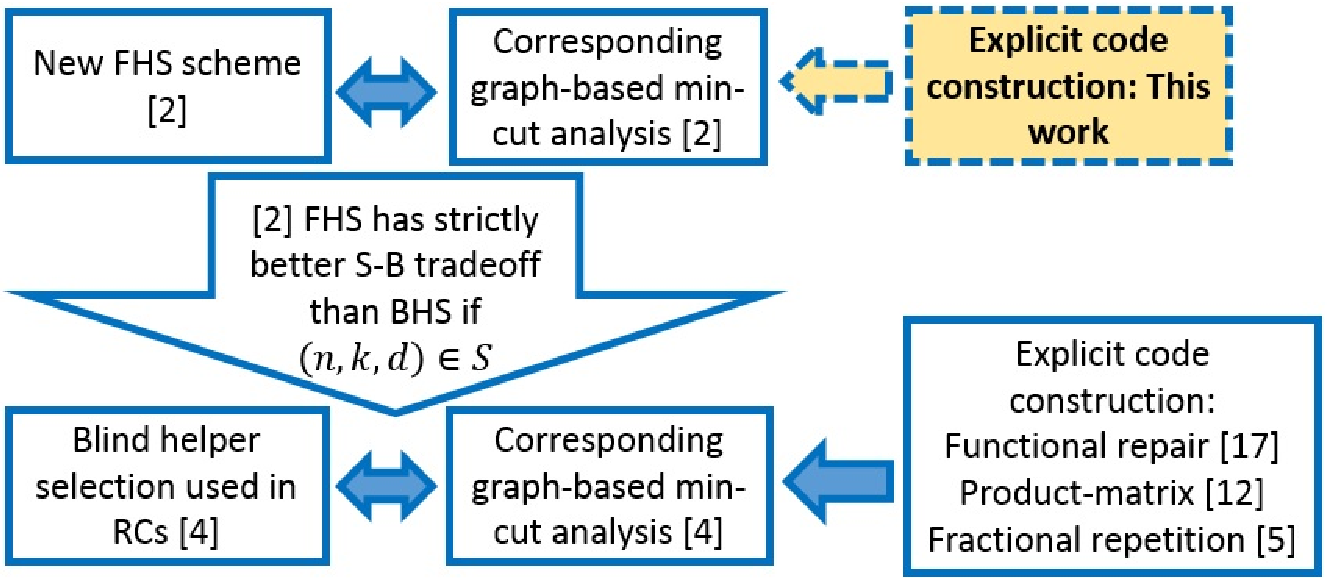}
\caption{Comparison of the DSN results: FHS is the achievability helper selection needed when characterizing the $(n,k,d)$ set $\nkdset$. This work discusses the pairing code construction.}
\label{tab}
\end{figure}

 Along a similar line of the functional-repair code construction in \cite{wu2010existence} and the subsequent exact-repair code constructions in \cite{rashmi2011optimal, el2010fractional}, this work focuses on explicit code design that can attain the purely graph-based S-B tradeoff curve of the FHS scheme \cite{ahmad2018can}. The results of this work thus complement the min-cut analysis of \cite{ahmad2018can} in the same way as \cite{wu2010existence} complements \cite{dimakis2010network}. Also see Fig.~\ref{tab} for the illustration.

\par The contribution of our new code construction, called \emph{flexible fractional repetition (FFR)}, is 3-fold: (i) it closes the loop of the min-cut-based analysis in \cite{ahmad2018can} by explicitly designing a code that achieves the MBR point of the FHS scheme, see Fig.~\ref{tab}. As a result, the fundamental question when can intelligent helper selection improve the S-B tradeoff is now answered with rigorous code construction, rather than the previous graph-based analysis.  (ii)  The proposed FFR construction can  be viewed as a generalization of the existing FR codes. However, unlike FR codes which require that $n\cdot d$ be even, our FFR codes are applicable to arbitrary $(n,k,d)$ values while retaining most of the practical benefits of FR codes, i.e., FFR codes are exact-repair codes and the majority of the nodes of a FFR code can be {\em repaired-by-transfer}.

\par (iii) The concept of the original FR codes is based on the ``repair-by-transfer (RBT) graph.''  An edge-counting argument for regular RBT graph is then developed in \cite{el2010fractional} and its subsequent works \cite{silberstein2015optimal,olmez2016fractional,koo2011scalable}. Our work develops an edge counting argument for {\em irregular} RBT graphs and also establishes the connection between the code-construction-based RBT graphs and the min-cut-based {\em information flow graphs} in \cite{ahmad2018can}. By matching the former (an achievability result) and the latter (a converse result), we prove the optimality of FFR under some $(n,k,d,\alpha,\beta)$, see Propositions~\ref{prop:optimal} and \ref{prop:optimal2}.  Such a new analysis approach will further enrich the literature of FR codes and broaden their applications.

\par The rest of this paper is organized as follows. Section~\ref{sec:related-work} reviews existing work on FR codes. Section~\ref{sec:flash} gives the notation used in this paper and states the existing results on FHS \cite{ahmad2018can}. Section~\ref{sec:examples} motivates FFR codes by providing two examples that demonstrate their construction. Section~\ref{sec:FFR} presents the construction of FFR codes. Section~\ref{sec:FFR_proof} sketches the analysis of FFR codes. Section~\ref{sec:conc} concludes the paper.

\section{Related Work on Fractional Repetition Codes} \label{sec:related-work}
\par The first construction of a special-case FR code appeared in \cite{rashmi2009explicit,shah2012distributed} (although not termed FR initially). The code construction in there was based on encoding the file first using an MDS code and then assigning the encoded packets to the edges of a complete graph. This code construction achieves the MBR point of RCs with $d=n-1$. Shortly after, the concept of assigning MDS-coded packets to edges of graphs was generalized to hypergraphs and the term ``FR codes'' was coined in \cite{el2010fractional}, in which the MDS code was referred to as the \emph{outer code} and the repetition code as the \emph{inner code}. 

\par In FR codes, the number of times packets are replicated in the network is termed the repetition degree. FR codes with repetition degree 2 were proposed in \cite{el2010fractional} based on regular graphs and are shown to achieve the MBR point of RCs with BHS if and only if $n\cdot d$ is even. Utilizing Steiner systems, \cite{el2010fractional} was able to construct FR codes for when the repetition degree is larger than 2. A subsequent work \cite{pawar2011dress} proposed DRESS codes that are randomly constructed codes utilizing the idea of FR codes. Reference \cite{koo2011scalable} presented graph-based constructions of Steiner systems that translate into FR codes for repetition degrees that are much smaller than the storage allowed per node. FR code constructions using {\em resolvable designs} were given in \cite{olmez2016fractional} that are able to cover a set of parameters not covered by Steiner systems. Moreover, \cite{zhu2014general} gave FR constructions for storage networks with heterogeneous numbers of helpers. 


\par All FR codes in the above mentioned works are based on the following two steps. Step~1: Construct MDS coded packets and duplicate each packet $r\geq 2$ times; and Step~2: the duplicated copies are carefully distributed and stored in network nodes. Those nodes that store the same packet will be helpers of each other when one of the them fails. Although the 2-step process is very powerful, only for a restricted collection of $(n,k,d)$ values can we successfully complete Steps~1 and 2, which thus limits the application of FR codes. 
\par In contrast, the main focus of the proposed FFR codes is not about designing Step~2. Instead, the helper selection of FFR codes is designed by the graph-based min-cut analysis in \cite{ahmad2018can}. Specifically, \cite{ahmad2018can} shows that in terms of the {\em min-cut values}, a new helper selection policy, called family helper selection (FHS), outperforms BHS whenever possible,\footnote{A more rigorous statement is: If there exists a helper selection that strictly outperforms BHS in terms of the S-B tradeoff, then FHS strictly outperforms BHS.} and is guaranteed to be {\em optimal} for some $(n,k,d)$ values. FFR codes directly use the FHS scheme in its Step~2 without any modification. However, it turns out that, with FHS in Step~2, it becomes impossible to reuse the original Step~1. Instead, the focus of the FFR codes is to modify Step~1 so that the combination of {\em the modified Step~1} and {\em the use of the FHS scheme in Step~2} results in a code that realizes the superior performance promised by the min-cut analysis in \cite{ahmad2018can}. By jointly revising Steps~1 and 2, the proposed FFR code can be applied to any $(n,k,d)$ values while retaining most of the practical appeals of the original FR codes, e.g., exact-repair and repair-by-transfer. This was previously not possible when the design efforts were focused on Step~2 only.

\section{Flashback of \cite{ahmad2018can} and Notation}\label{sec:flash}
\par We denote the total number of nodes in a network by $n$. The number of helper nodes, the nodes participating in the repair of a failed node, is denoted by $d$. This means that during repair, the node that is replacing the failed node, called the newcomer, can contact $d$ nodes for repair.  For the reliability requirement, we require that any set of $k$ nodes of the total $n$ nodes must\footnote{For a detailed explanation of the parameters $d$ and $k$ and the distinction between the \emph{desired protection level} $k$ and the \emph{actual achievable protection level} $k^*$ of a code, see \cite{ahmad2018can}.} be able to reconstruct the original data. 

\par The performance of a system is measured by the amount of storage-per-node, $\alpha$,  the amount of communications or bandwidth-per-helper, $\beta$, and the size of the original data/file, $\mathcal{M}$. See \cite{ahmad2018can,dimakis2010network} for detailed definitions of $(n,k,d,\alpha,\beta,\mathcal{M})$. 

\par An {\em intelligent} helper selection scheme chooses the $d$ helpers carefully based on the past failure patterns. In contrast, a blind helper selection (BHS) scheme chooses the $d$ helpers {\em blindly}. We then have

\begin{definition}
An arbitrarily given $(n,k,d)$ value is {\em indifferent-to-helper-selection} (ITHS) if there exists no intelligent helper selection scheme (even with unlimited computing power) that can strictly outperform BHS in terms of the achievable $(\alpha,\beta,\mathcal{M})$. 
\end{definition}

For example, if the $(n,k,d)$ satisfies $d=n-1$, then such $(n,k,d)$ is clearly ITHS. The reason is that when $d=n-1$, there is only one way of choosing the $d=n-1$ helpers, i.e., all the remaining $n-1$ nodes must help the newcomer. Therefore, there is no room for improvement for any intelligent helper selection scheme and the $(n,k,d)$ is clearly ITHS by definition. Surprisingly, there are many other $(n,k,d)$ values with $d<n-1$ that are also ITHS. That is, even if we have the new degree of freedom of carefully choosing $d$ out of $n-1$ remaining nodes, for some $(n,k,d)$ no additional performance can be gained by intelligent helper selection. 

\par Knowing whether $(n,k,d)$ is ITHS or not is very beneficial. For example, if a given $(n,k,d)$ is not ITHS, then there must exist an intelligent helper selection scheme that can strictly outperform BHS and the network designer should focus on how to harvest the promised performance gain. \cite{ahmad2018can} finds the following necessary and sufficient condition for all ITHS $(n,k,d)$ values.

\begin{proposition}\cite[Propositions~1 and 2]{ahmad2018can}\label{prop:comparison_new} An $(n,k,d)$ value is ITHS if and only if at least one of the following two conditions is true (i) $d=1$, $k=3$, and $n$ is odd; and (ii) $k\leq \left\lceil \frac{n}{n-d}\right\rceil$.
\end{proposition}

\par The achievability part of the above results (the only if direction), i.e., proving the existence of a helper selection scheme that can do strictly better, is established by analyzing the min-cut values of a new class of helper selection schemes termed \emph{the family helper selection (FHS) scheme}. The basic concepts and notation of the FHS scheme will be introduced in the next subsection.

\subsection{The Family Helper Selection Scheme}\label{sec:fhs}
At any time $t$, the helper choices of the FHS scheme are described by $\{D_i: i=1,\cdots, n\}$, where $D_i$ is the helper set when the $i$-th node fails and contains exactly $d$ nodes. The sets $\{D_i\}$ are defined as follows. Label the storage nodes by $1$ to $n$. Then, the first $(n-d)$ nodes are grouped as the first \emph{complete family} and the second $(n-d)$ nodes are grouped as the second complete family and so on. In total, there are $c\triangleq\left\lfloor \frac{n}{n-d}\right\rfloor$ complete families. The remaining $n\bmod(n-d)$ nodes, if there are any, are grouped as an {\em incomplete family}. For any node $i$ in a complete family, the helper set $D_i$ contains all the nodes {\em not} in the same family of node $i$. For any node $i$ in the incomplete family, we choose $D_i=\{1,\cdots, d\}$, the first $d$ nodes. An example of the FHS scheme will be provided in Section~\ref{sec:ex3}. 

\section{Two Examples That Demonstrate the Construction of FFR Codes}\label{sec:examples}

We now present two examples demonstrating the difference between FR and our new proposed FFR codes. 

\subsection{Example 1: Not All FR Codes Are Equal}
\par
\begin{figure}[h!]
\centering
\includegraphics[width=0.37\textwidth]{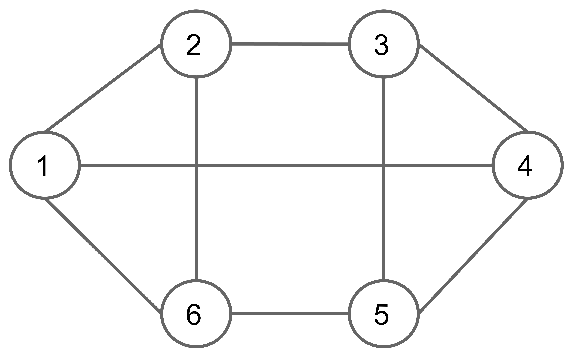}
\caption{A regular graph of the FR code for $(n,k,d,\alpha,\beta)=(6,3,3,3,1)$.}
\label{fig:ex2}
\end{figure}

 Consider the parameter value $(n,k,d,\alpha,\beta)=(6,3,3,3,1)$. As described in \cite{el2010fractional}, the FR code is based on finding a regular graph of $n=6$ nodes and node degree $d=3$. This is possible since $n\cdot d$ is even. One (possible) regular graph for this $(n,k,d,\alpha,\beta)$ is shown in Fig.~\ref{fig:ex2}.

Using the regular graph in Fig.~\ref{fig:ex2}, we can construct an FR code that can protect $\mathcal{M}=6$ packets. The construction is as follows. First use a $(9,6)$-MDS code to convert the $\mathcal{M}=6$ original packets into 9 MDS-coded packets. Then, each MDS coded packet is assigned to one of the 9 edges in Fig.~\ref{fig:ex2}. Each node will then store the $d=3$ packets corresponding to its 3 adjacent edges. To see that any $k=3$ nodes can reconstruct the original file, we observe\footnote{This can be verified by a simple computer program that examines all ${6\choose 3}$ node combinations and counts the adjacent edges.} that any nodes have $\geq 6$ distinct edges incident to them. E.g., nodes $\{1,2,6\}$ have exactly 6 adjacent edges.  Then, by the MDS property, these $\geq 6$ MDS-coded packets can be used to reconstruct the original file. 

\par Note that \cite{dimakis2010network} proves that if BHS is used, then a distributed storage network with $(n,k,d,\alpha,\beta)=(6,3,3,1,1)$ can protect at most $\mathcal{M}=6$. However, by plugging in the $(n,k,d)=(6,3,3)$ value into Proposition~\ref{prop:comparison_new}, we have  $k=3>\left\lceil\frac{n}{n-d}\right\rceil=\left\lceil\frac{6}{3}\right\rceil=2$, which implies that there exists an intelligent helper selection scheme that strictly outperforms the best performance of BHS \cite{dimakis2010network}, which is $\mathcal{M}=6$. This thus prompts the question whether we can design an FR code of $(n,k,d,\alpha,\beta)=(6,3,3,1,1)$ that can protect a larger file, say $\mathcal{M}=7$.

\par
\begin{figure}[h!]
\centering
\includegraphics[width=0.23\textwidth]{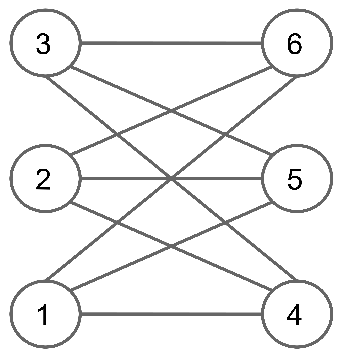}
\caption{An alternative regular graph for $(n,k,d,\alpha,\beta)=(6,3,3,3,1)$.}
\label{fig:ex2_2}
\end{figure}

 We observe that the regular graph for this example is actually not unique. Instead, we can consider another regular graph in Fig.~\ref{fig:ex2_2} which also has $n=6$ nodes and node degree $d=3$. We observe that in this new regular graph, any $k=3$ nodes have $\geq 7$ distinct adjacent edges.\footnote{Again a simple computer program can verify this fact.} As a result, if we use a $(9,7)$-MDS code in the beginning and use the regular graph in Fig.~\ref{fig:ex2_2}, then the resulting FR code can protect a file of size $\mathcal{M}=7$.

\par This example demonstrates that the performance of an FR code depends on how one chooses the underlying regular graph. Perhaps more importantly, it hints that the helper selection benefits promised by the min-cut analysis in Proposition~\ref{prop:comparison_new} can be realized by a clever construction of FR codes, at least for the case of $(n,k,d,\alpha,\beta)=(6,3,3,1,1)$.  Our proposed FFR codes build on top of these two observations.  That is, we generalize FR codes for irregular graphs and show that it is true that any helper selection benefits promised by the min-cut analysis in Proposition~\ref{prop:comparison_new} can indeed by realized by our FFR codes.


\subsection{Example 2: Sometimes No FR Code Is Good Enough} \label{sec:ex3}
\par We use the parameter value $(n,k,d,\alpha,\beta)=(7,3,3,3,1)$ to demonstrate the limitation of FR codes and how our FFR codes work. For $(n,k,d,\alpha,\beta)=(7,3,3,3,1)$, \cite{dimakis2010network} proves that BHS can protect a file of size $\mathcal{M}=6$. Again by plugging in Proposition~\ref{prop:comparison_new}, we have $k=3>\left\lceil\frac{n}{n-d}\right\rceil=\left\lceil\frac{7}{4}\right\rceil=2$, which implies that there exists an intelligent helper selection scheme that can protect $\mathcal{M}>6$ packets, say protect $\mathcal{M}=7$ packets. The remaining question is how to design such a scheme.

\par Following the success in Example~1, one may like to directly apply the FR code in this scenario. However, for this $(n,k,d)=(7,3,3)$ it is provably impossible to find any regular graph with $n=7$ nodes and degree $d=3$.  Therefore, no FR code exists for $(n,k,d)=(7,3,3)$. In the following, we show how our FFR code works for $(n,k,d)=(7,3,3)$. 

\par
\begin{figure}[h!]
\centering
\includegraphics[width=0.45\textwidth]{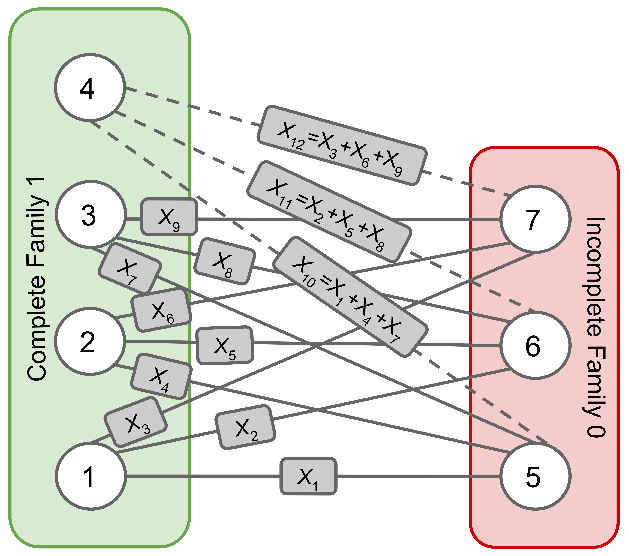}
\caption{The graph representation of the code for $(n,k,d,\alpha,\beta)=(7,3,3,3,1)$.}
\label{fig:ex3}
\end{figure}

 First, we have that $\left\lfloor\frac{n}{n-d}\right\rfloor=\left\lfloor\frac{7}{4}\right\rfloor=1$. Following the FHS description in Section~\ref{sec:fhs}, we have 1 complete family, nodes $\{1,2,3,4\}$, and 1 incomplete family, nodes $\{5,6,7\}$. More specifically, any of nodes 1 to 4 will request help from nodes 5 to 7. Any of nodes 5 to 7 will request help from nodes 1 to 3. Note the asymmetry of the helper relationship, i.e., node 4 requests help from nodes 5 to 7 but is not a helper for any of nodes 5 to 7. See Fig.~\ref{fig:ex3} for illustration, in which we use the dashed line to represent the asymmetric helper relationship of node 4.

\par Our FFR code is based on GF$(32)$ and can protect a file of 7 packets while satisfying $(n,k,d,\alpha,\beta)=(7,3,3,3,1)$. The 7 packets of the file are denoted by $W_1,W_2,\dots,W_7$. We first encode the 7 packets into 9 packets $X_1$ to $X_9$ where $X_i=W_i$ for $i=1$ to 7 and $X_8$ and $X_9$ are
\begin{align}
X_8=23W_1+&3W_2+9W_3+24W_4+\nonumber\\
&\quad \quad \quad \quad 30W_5+8W_6+8W_7, \label{eq:phase1-p2-eq1} \\
X_9=25W_1+&25W_2+2W_3+18W_4+\nonumber\\
&\quad \quad \quad \quad 12W_5+25W_6+27W_7. \label{eq:phase1-p2-eq2}
\end{align}
Finally, we create 3 additional packets $X_{10}$, $X_{11}$, and $X_{12}$ by
\begin{align}
& X_{10}=X_1+X_4+X_7, \label{eq:phase1-p1-eq1} \\
& X_{11}=X_2+X_5+X_8, \label{eq:phase1-p1-eq2} \\
& X_{12}=X_3+X_6+X_9. \label{eq:phase1-p1-eq3}
\end{align}
Once the $X_1$ to $X_{12}$ packets are encoded from $W_1$ to $W_7$, we assign the packets $X_1,X_2,\dots,X_9$ to the solid edges as shown in Fig.~\ref{fig:ex3} and assign the packets $X_{10}, X_{11},$ and $X_{12}$ to the dashed edges incident to incomplete family nodes 5, 6, and 7, respectively. Each physical node in $\{1,2,3,5,6,7\}$ (excluding node 4) stores the packets corresponding to the \emph{solid} edges adjacent to it. Node 4 stores the packets corresponding to the dashed edges incident to itself.\footnote{The FFR code construction for general $(n,k,d)$ values will be detailed in Section~\ref{sec:FFR_des}.} One can clearly see that, in this code construction, each node stores exactly $\alpha=3$ packets.

\par \textbf{Repair:} If any of the nodes in $\{1,2,3,5,6,7\}$ (excluding node 4) fails, then the newcomer downloads the lost packets of the solid edges from its adjacent nodes. If node 4 fails, then nodes 5, 6, and 7 generate and send to the newcomer the linear combinations $X_1+X_4+X_7$, $X_2+X_5+X_8$, and $X_3+X_6+X_9$, respectively. This is always possible since node 5 stores $\{X_1,X_4,X_7\}$, node 6 stores $\{X_2,X_5,X_8\}$, and node 7 stores $\{X_3,X_6,X_9\}$. Notice that these generated packets correspond to the packets $X_{10},X_{11},$ and $X_{12}$ of the dashed edges, see \eqref{eq:phase1-p2-eq1} and \eqref{eq:phase1-p2-eq2}, and node 4 is thus exactly-repaired. Our FFR construction is {\em almost repairable-by-transfer}, since all nodes but node 4 can be repaired by transfer.

\par \textbf{Reconstruction:} One can verify, by a computer-based exhaustive search, that the given code assignment can reconstruct the $\mathcal{M}=7$ packets of the original file from any $k=3$ nodes of the total $n=7$ nodes. That is, we use a computer to verify that the coding matrix of the packets in any $k=3$ nodes is always of full rank 7.  Note that in FFR, one cannot simply count the edges as in \cite{el2010fractional}. Instead one has to check for the matrix invertability since the underlying graph, see Fig.~\ref{fig:ex3}, is non-regular and of asymmetric helper relationship (solid versus dashed edges). One of the contributions of this work is to analytically characterize the protected file size $\mathcal{M}$ of our FFR codes for arbitrary $(n,k,d)$ values.

\section{The Flexible Fractional Repetition Codes}\label{sec:FFR}
The motivation of the FFR design is to achieve the MBR point of the FHS scheme computed by the min-cut analysis in \cite{ahmad2018can}.  See Fig.~\ref{tab}.  In this section, we will first describe the MBR point of the FHS scheme and then describe the FFR construction that attains it.

\subsection{The MBR Point of the FHS Scheme}

\def\nfam{ n_\text{fam}}
Define $\nfam=\left\lceil \frac{n}{n-d}\right\rceil$ and define a sequence of $n$ numbers $w_1$ to $w_n$ by
\begin{align}
&(w_1,\cdots, w_n)=\nonumber\\
&\left(\overbrace{0,\cdots, 0}^{\nfam}, \overbrace{1,\cdots, 1}^{\nfam}, \cdots, \overbrace{ \delta, \cdots, \delta}^{\nfam},\overbrace{ \delta+1, \cdots,\delta+1}^{n\bmod \nfam} \right),\label{eq:direct_y_index}
\end{align}
where $\delta\triangleq \left\lfloor \frac{n}{\nfam}\right\rfloor-1$.  Namely, $w_i$ contains a strictly increasing integer sequence $0, 1, 2,  \cdots$ with each entry repeated for $\nfam$ times. The value $\delta$ is the last entry that can be repeated for $\nfam$ times. The values of the remaining $(n\bmod \nfam)$ entries are assigned to $\delta+1$. With the above construction of $w_i$, we define $y_i\triangleq(i-1)-w_i$ for all $i=1$ to $n$.
\begin{proposition}\cite[Proposition~6]{ahmad2018can}\label{prop:mbr}
For any given $(n,k,d,\alpha,\beta)$ values satisfying $\alpha=d\beta$, thus the MBR point, the largest file size $\mathcal{M}$ that can be protected by the FHS scheme is
\begin{align}
\mathcal{M}=\sum_{i=1}^{k} (d-y_i)\beta \label{eq:mbr}.
\end{align}
\end{proposition}

\par For example, if $(n,k,d,\alpha,\beta)=(7,3,3,3,1)$, then we have $\nfam=\lceil 7/(7-3)\rceil=2, (w_1,w_2,w_3)=(0, 0, 1)$, and $(y_1,y_2,y_3)=(0,1,1)$. The protected file size $\mathcal{M}$ becomes $\sum_{i=1}^3(3-y_i)=(3-0)+(3-1)+(3-1)=7$.

\par Also note that Proposition~\ref{prop:mbr} characterizes the performance of FHS by a pure min-cut-based analysis. In Proposition~\ref{prop:FFR_rec} of Section~\ref{subsec:rec}, we prove that the MBR point described in Proposition~\ref{prop:mbr} can be achieved by our FFR code construction for any $(n,k,d)$ values.

\subsection{The Construction of FFR Codes}\label{sec:FFR_des}
\par Before describing the construction of FFR codes, we list some notational definitions. Consider the FHS scheme described in Section~\ref{sec:fhs}, we denote the set of nodes of the $i$-th complete family by $N_i$. Recall that there are $ c \triangleq\left\lfloor \frac{n}{n-d}\right\rfloor$ complete families. For the last complete family, i.e., $i=c$, we split its nodes into two disjoint node sets, $N_{-c}$ is the set of nodes in family $c$ that is {\em not} in the helper set of the incomplete family nodes and $N_c$ is the set of the remaining nodes of this complete family.  We denote the set of nodes in the incomplete family by $N_0$. The set of all nodes in the network is denoted by $N$. For example, if $(n,d)=(7,3)$ as in the example of Section~\ref{sec:ex3}, we have $c=1$ complete family, $N_1=\{1,2,3\}$, $N_{-1}=\{4\}$, and $N_0=\{5,6,7\}$. 

\par In short, we denote the incomplete family as family $0$, and split the last complete family, family $c$, into two family indices $c$ and $-c$, where the latter represents those nodes that are not helpers of any node. See Fig.~\ref{fig:ex3}. Then, $N_x$ contains the nodes that have family index $x$. For any node $i\in \mathcal{N}_x$, we define the inverse map $x=FI(i)$, which stands for the {\em family index} of $i$. In the above example, $FI(i)=1$ for $i=1$ to $3$, $FI(4)=-1$, and $FI(i)=0$ for $i=5$ to $7$. 

\par We assume without loss of generality that $\beta=1$ and $\alpha=d$ with the unit being ``packets''. The goal of FFR codes is to protect a file of size described in \eqref{eq:mbr} against any $(n-k)$ simultaneous failures. Since $\beta=1$, we can rewrite \eqref{eq:mbr} by 
\begin{align} \label{eq:file_size_construction}
\mathcal{M}=\sum_{i=1}^{k} \left(d-y_i\right) \text { packets.}
\end{align}
In all the subsequent discussions, we assume $\mathcal{M}$ is a fixed integer computed by \eqref{eq:file_size_construction}. 

\par The core idea of FFR codes stems from the concatenation of an inner code that is based on a graph representation of the distributed storage network and a carefully designed outer code that satisfies special properties. We first introduce the graph-based inner code of the FFR code.

\par {\bf The inner code:} The inner code is based on the following graph representation of the distributed storage network. Each physical node in the network is represented by a vertex in the graph, which is denoted by $G=(V,E)$ where $V$ denotes the set of vertices of $G$ and $E$ denotes its set of edges. As will be described, the graph consists of two disjoint groups of edges. Graph $G$ has the following properties:

\begin{enumerate}
\item $V=\{1,2,\cdots, n\}$. Each vertex $i$ in $V$ corresponds to physical node $i$ in $N$. For convenience, throughout our discussion, we simply say vertex $i\in N_x$ if the physical node that vertex $i$ corresponds to is in $N_x$.
\item Any two vertices $i\in N_x$ and $j\in N_y$ are connected by an edge in $E$ if $|x|\neq|y|$ and $(x,y)\notin \{(0,-c),(-c,0)\}$. The collection of all those edges is denoted by $\bar{E}$.
\item Any two vertices $i\in N_0$ and $j\in N_{-c}$ are connected by an edge in $E$. The collection of all those edges is denoted by $\tilde{E}$.
\item From the above construction, we have $E=\bar{E}\cup\tilde{E}$. We further assume that all the edges are undirected.
\end{enumerate}

Fig.~\ref{fig:ex3} of Example~2 in Section~\ref{sec:examples} is an example of the above graph representation of the inner code. Notice that the edges in $\bar{E}$ are represented by solid lines, while the edges in $\tilde{E}$ are represented by dashed lines.


\par Recall that $FI(i)$ denotes the family index of node $i$. We define the following three sets:
\begin{align}
{\sf IJ}^{[1]}&= \{(i,j): 1\leq i< j\leq n, 1\leq|FI(i)|<|FI(j)| \leq c\}\nonumber\\
{\sf IJ}^{[2]}&= \{(i,j): 1\leq i<j \leq n, 1\leq FI(i)\leq c,FI(j)=0\}\nonumber\\
{\sf IJ}^{[3]}&= \{(i,j): 1\leq j<i \leq n, FI(i)=0, FI(j)=-c\}\nonumber.
\end{align}

One can easily verify that the union of the first two sets, ${\sf IJ}^{[1]}\cup{\sf IJ}^{[2]}$, can be mapped bijectively to the edge set $\bar{E}$, and the third set ${\sf IJ}^{[3]}$ can be mapped bijectively to the edge set $\tilde{E}$. The difference between sets ${\sf IJ}^{[1]}$, ${\sf IJ}^{[2]}$ and ${\sf IJ}^{[3]}$ and $\bar{E}$ and $\tilde{E}$ is that the sets ${\sf IJ}^{[1]}$ to ${\sf IJ}^{[3]}$ focus on {\em ordered pairs} while the edges in $E$ correspond to unordered vertex pairs (undirected edges). 

\par The unordered edge sets $\bar{E}$ and $\tilde{E}$ capture the main design ideas in a more intuitive way while the ordered sets ${\sf IJ}^{[1]}$ to ${\sf IJ}^{[3]}$ are easier to use during the actual counting process. For example,  there are $\frac{(n-|N_0|)(d-|N_0|)}{2}$ pairs in ${\sf IJ}^{[1]}$, $d|N_0|$ pairs in ${\sf IJ}^{[2]}$, and $|N_{-c}|\cdot|N_0|$ pairs in ${\sf IJ}^{[3]}$. Thus, in total, there are
\begin{align}\label{eq:total_packets}
\frac{(n-|N_0|)(d-|N_0|)}{2}+d|N_0|+|N_{-c}|\cdot |N_0|
\end{align}
distinct pairs in the overall index set ${\sf IJ}^{[1]}\cup {\sf IJ}^{[2]}\cup {\sf IJ}^{[3]}$. This implies that the total number of edges of graph $G$ is also characterized by \eqref{eq:total_packets}.
\par Each edge of graph $G$ corresponds to one coded packet that is stored in the distributed storage system. More specifically, each edge $(i,j)\in \bar{E}$ represents a packet $P_{(i,j)}$ that is stored in the two physical nodes $i$ and $j$, i.e., \emph{both} nodes $i$ and $j$ store an identical copy of the packet $P_{(i,j)}$. On the other hand, each edge $(i,j)\in\tilde{E}$ represents a packet $\tilde{P}_{(i,j)}$ that is stored in \emph{only one} of its two vertices, the corresponding vertex in $N_{-c}$. One can verify by examining the ${\sf IJ}^{[1]}$ to ${\sf IJ}^{[3]}$ index sets defined previously that each physical node stores exactly $\alpha=d$ packets.

\par {\bf The outer code:} We now describe how to generate the $|{\sf IJ}^{[1]}|+|{\sf IJ}^{[2]}|+|{\sf IJ}^{[3]}|$ coded packets (the $P_{(i,j)}$ and $\tilde{P}_{(i,j)}$ packets depending on whether $(i,j)\in\bar{E}$ or $(i,j)\in\tilde{E}$) from the $\mathcal{M}$ original packets, where $\mathcal{M}$ is specified by \eqref{eq:file_size_construction}. Our goal is to design the $|{\sf IJ}^{[1]}|+|{\sf IJ}^{[2]}|+|{\sf IJ}^{[3]}|$ coded packets satisfying  the following two properties.

\par {\bf Property~1:} For any $i_0\in N_0$, there are $d$ different $j$ indices satisfying $(j,i_0)\in {\sf IJ}^{[2]}$ and they are those $j\in N_1\cup N_2\cup \cdots \cup N_c$ for all $(j,i_0)\in {\sf IJ}^{[2]}$. We require that any given coded packet $\tilde{P}_{(i_0,j)}$ corresponding to some $(i_0,j)\in {\sf IJ}^{[3]}$ must be a linear combination of the $d$ packets $P_{(j_0,i_0)}$ for all $j_0$ satisfying $(j_0,i_0)\in {\sf IJ}^{[2]}$, i.e., those $d$ packets stored in node $i_0$. 

\par We now describe the second required property. Recall that there are $|N_0|=n\bmod (n-d)$ nodes in the incomplete family and they have {\em node} indices $c(n-d)+1$ to $c(n-d)+|N_0|$ where $c$ is the family index of the last complete family. Consider any arbitrary but fixed subset of edges $E_\text{sub}\subseteq \bar{E}\cup\tilde{E}$ and we will define $(|N_0|+1)$ different values $a_0$ to $a_{|N_0|}$ in the following way. Define $a_m$, $m=1$ to $|N_0|$, as the number of edges $e\in E_\text{sub}$ satisfying that $e$ is connected to the node $(c(n-d)+m)$, the  $m$-th vertex in $N_0$. By definition, it is clear that $\sum_{m=1}^{|N_0|} a_m=|E_{\text{sub}}\cap({\sf IJ}^{[2]}\cup {\sf IJ}^{[3]})|$, where we abuse the notation slightly by treating the ordered-pair sets ${\sf IJ}^{[2]}$ and ${\sf IJ}^{[3]}$ as unordered edge sets. Define $a_0$ as the number of $e\in E_{\text{sub}}$ that are not connected to any of the vertices in $N_0$, i.e., $a_0\triangleq|E_{\text{sub}}\cap {\sf IJ}^{[1]}|$. Define $\mathsf{a.count}\stackrel{\Delta}{=} a_0+\sum_{m=1}^{|N_0|}\min(a_m,d)$. The above description specifies how to compute a value $\mathsf{a.count}$ from any given $E_{\text{sub}}$. 

\par The intuition of this $\mathsf{a.count}$ computation is as follows. In the traditional FR construction, each edge carries a distinct packet generated by an MDS code. Therefore the packets are as linearly independent as possible. The rank of the corresponding coding matrix is thus equal to the number of distinct edges. However, the construction of FFR has to satisfy Property~1. That is, the packets corresponding to those edges in $\mathsf{IJ}^{[3]}$ must be a linear sum of the $\alpha=d$ packets stored in the node $i_0\in N_0$, see Property~1. Therefore, the packets are not as independent to the same degree as with the MDS-code-based construction. As a result, one uses the minimum between $a_m$ and $d$ to take into account the linear dependency imposed by Property~1. The value of $\mathsf{a.count}$ then represents an upper bound on the rank of the coding matrix corresponding to the packets in edges $E_\text{sub}$. The following Property~2 then imposes that the packets must be as linearly independent as possible, with the matrix rank meeting the upper bound $\mathsf{a.count}$ whenever possible.

\par {\bf Property~2:} The $|E|$ coded packets must satisfy that for any subset of edges $E_\text{sub}$ satisfying $\mathsf{a.count} \geq \mathcal{M}$, the corresponding packets  can be used to reconstruct the original file of size $\mathcal{M}$ packets, i.e., the corresponding coding matrix being of full rank. 

\par  In the following, we describe how to construct  the outer code, i.e., how to design coded packets for the $|E|$ edges that satisfy the above two properties. Specifically, we can use a two-phase approach to generate the packets. We first independently and uniformly randomly generate $|\bar{E}|=\frac{(n-|N_0|)(d-|N_0|)}{2}+d|N_0|$ linearly encoded packets from the $\mathcal{M}$ packets of the original file. These packets are fixed and arbitrarily assigned to the edges in $\bar{E}$ (one for each edge). After this first step, all physical nodes store exactly $d$ packets except those nodes in $N_{-c}$, each of which now stores exactly $(d-|N_0|)$ packets. Now, from each node in $u\in N_0$, we generate independently and uniformly a random set of $|N_{-c}|$ linearly encoded packets from the $d$ packets stored in $u$. We fix these newly generated packets and assign them to each of the $|N_{-c}|$ edges in $\{(u,w)\in \tilde{E}:\forall w\in N_{-c}\}$. Specifically, these $|N_{-c}|$ packets will now be stored in node $w\in N_{-c}$, one for each $w$. Repeat this construction for all $u\in N_0$. After this second step, each edge in $\bar{E}\cup \tilde{E}$ has been assigned one distinct coded packet and each node in $N=N_1\cup\cdots N_{c}\cup N_{-c}\cup N_0$ now stores exactly $d$ packets. The Phase~1 construction is now complete.

\par After the initial random-construction phase, we enter the second phase, the verification phase. In this phase, we fix the packets and deterministically check whether they satisfy Property~2 (by our construction the coded packets always satisfy Property~1). The following lemma states that with high probability, the randomly generated packets in Phase~1 will satisfy Property~2.

\begin{lemma}\label{lem:FFR_existence}
When $\text{GF}(q)$ is large enough, with close-to-one probability, the above random construction will satisfy Property~2.
\end{lemma}

\par The proof of Lemma~\ref{lem:FFR_existence} is relegated to Appendix~\ref{app:FFR_existence}.


\par To illustrate the construction/notation of FFR codes, we return to Example~2 of Section~\ref{sec:examples}. In that example, we have $(n,k,d,\alpha,\beta)=(7,3,3,3,1)$ and $|E|=12$, $|\bar{E}|=9$, and $|\tilde{E}|=3$, see Fig.~\ref{fig:ex3}. The packets corresponding to the edges in $\bar{E}$ are $P_{(1,5)}=X_1$, $P_{(1,6)}=X_2$, $P_{(1,7)}=X_3$, $P_{(2,5)}=X_4$, $P_{(2,6)}=X_5$, $P_{(2,7)}=X_6$, $P_{(3,5)}=X_7$, $P_{(3,6)}=X_8$, and $P_{(3,7)}=X_9$. On the other hand, the packets corresponding to edges in $\tilde{E}$ are $\tilde{P}_{(5,4)}=X_{10}$, $\tilde{P}_{(6,4)}=X_{11}$, and $\tilde{P}_{(7,4)}=X_{12}$. It is clear by \eqref{eq:phase1-p1-eq1} to \eqref{eq:phase1-p1-eq3} that the construction satisfies Property~1. The coefficients in \eqref{eq:phase1-p2-eq1} and \eqref{eq:phase1-p2-eq2} are chosen randomly while using computers to verify that Property~2 is satisfied for the final construction.

\section{Analysis of FFR Codes}\label{sec:FFR_proof}

\subsection{The Repair Operations}
In this section, we first argue that FFR codes can be exactly-repaired using FHS. First, consider the case that node $i$ fails for some $i\in N_1\cup N_2\cup \cdots\cup N_c\cup N_0$ (those in $N\backslash N_{-c}$). The $d$ packets stored in node $i$ thus need to be repaired. We then notice that the $d$ packets in node $i$ correspond to the $d$ edges in $\bar{E}$ that are incident to node $i$. Therefore, each of those $d$ packets to be repaired is stored in another node $j$ and node $i$ can thus be {\em repaired-by-transfer}. Note that by our construction, the neighbors of node $i$ are indeed the helper set $D_i$ in FHS. Also see our discussion in Example~2 of Section~\ref{sec:examples} for illustration.

\par We now consider the case in which node $i$ in $N_{-c}$ fails. We again notice that $(c-1)(n-d)$ of its $d$ packets correspond to (solid) edges in $\bar{E}$. Therefore, each of those $(c-1)(n-d)$ packets is also stored in another node and can again be {\em repaired-by-transfer}. To restore the remaining $n\bmod(n-d)$ packets, by our construction, these packets correspond to the edges in $\{(w,i)\in \tilde{E}:(w,i)\in {\sf IJ}^{[3]}\}$. By Property~1 of our outer code construction, for any $w_0\in N_0$, $\tilde{P}_{(w_0,i)}$, of those in ${\sf IJ}^{[3]}$, is a linear combination of the $d$ packets $\{P_{(j,w_0)}:( j,w_0)\in {\sf IJ}^{[2]},  j=1,2,\cdots, d\}$ stored in node $w_0$. Thus, during repair, newcomer $i$ can ask physical node $w_0$ to compute the packet $\tilde{P}_{(w_0,i)}$ and send the final result for all $w_0\in N_0$. Therefore, newcomer $i$ can exactly-repair all the remaining $n\bmod(n-d)$ packets as well. Also see our discussion in Example~2 of Section~\ref{sec:examples} for illustration.

\subsection{The Reconstruction Operations}\label{subsec:rec}

\par The following proposition shows that the FFR code with FHS can protect against any $(n-k)$ simultaneous failures.
\begin{proposition}\label{prop:FFR_rec}
Consider the FFR code with any given $(n,k,d)$ values. For any arbitrary selection of $k$ nodes, one can use all the $k\cdot d$ packets stored in these $k$ nodes (some of them may be identical copies of each other) to reconstruct the original file $\mathcal{M}$ packets with $\mathcal{M}$ described in \eqref{eq:file_size_construction}.
\end{proposition}

The proof of Proposition~\ref{prop:FFR_rec} will be provided shortly. Since the $\alpha$, $\beta$, and $\mathcal{M}$ values in \eqref{eq:file_size_construction} match the MBR point of the FHS scheme in \eqref{eq:mbr}, Proposition~\ref{prop:FFR_rec} shows that the explicitly constructed FFR code indeed achieves the MBR point of the FHS scheme predicted by the min-cut-based analysis. It turns out that FFR is indeed optimal in some scenarios.

\begin{proposition} \label{prop:optimal}
If $d$ is even, $n=d+2$, $k=n/2+1$, and $\alpha=d\beta$, then FFR is optimal. Namely, there is no code that can protect a file size strictly larger than the protected file size of FFR characterized by \eqref{eq:mbr}.
\end{proposition}

\begin{proposition} \label{prop:optimal2}
If $n\mod(n-d)=0$, $k=n-1$, and $\alpha=d\beta$, then FFR is again optimal, i.e., achieves the largest $\mathcal{M}$. 
\end{proposition}

Propositions~\ref{prop:optimal} and \ref{prop:optimal2} are direct corollaries of Proposition~\ref{prop:FFR_rec} and \cite[Propositions~3 and 4]{ahmad2018can}. Their proofs are thus omitted. 

\par The rest of this subsection is dedicated to the proof of Proposition~\ref{prop:FFR_rec}.
\begin{IEEEproof}
Consider an arbitrarily given set of $k$ nodes in the distributed storage network, denoted by $S$. Denote nodes in $S$ that belong to $N_i$ by $S_i\stackrel{\Delta}{=}S\cap N_i$. We now consider the set of edges that are incident to the given node set $S$, i.e., those edges have at least one end being in $S$ and each of the edges corresponds to a distinct packet stored in nodes $S$. Recall that for any set of edges, we can compute the corresponding $\mathsf{a.count}$ value as defined in Property~2 of our code construction. The key to the proof is to show that the $\mathsf{a.count}$ value of the edges incident to $S$ is no less than $\mathcal{M}$. Then Property~2 immediately leads to the proof of Proposition~\ref{prop:FFR_rec}. 

\par To that end, we describe the following step-by-step procedure, termed {\sc Count}, that computes the value $\mathsf{a.count}$. We will later analyze each step of the procedure to quantify the $\mathsf{a.count}$ value.

\begin{enumerate}
\item We first define $G_1=(V_1,E_1)=G=(V,E)$ as the original graph representation of the FFR code. Choose an arbitrary order for the vertices in $S$ such that all nodes in $S_{-c}$ come last. Call the $i$-th vertex in the order, $v_i$. That is, we have that $S_{-c}=\{v_i:k-|S_{-c}|+1\leq i\leq k\}$ and $S_1\cup\cdots\cup S_c\cup S_0=\{v_i:1\leq i\leq k-|S_{-c}| \}$.
\item Set $e(S)=0$, where $e(S)$ will be used to compute $\mathsf{a.count}$.

Now, do the following step sequentially for $i=1$ to $|S|=k$:
\item Consider vertex $v_i$. We first compute
\begin{align}\label{eq:inter9}
x_i&=|\{(v_i,j)\in E_i \cap \bar{E}:j\in N\}|+1_{\{v_i\in S_{-c}\}}\cdot\nonumber\\
& \quad \quad \sum_{u\in N_0} 1_{\{(u,v_i) \in E_i\cap \tilde{E}\}}\cdot 1_{\{|\{(u,j)\in E_i :j\in N\}|>|N_{-c}| \}}.
\end{align}
Once $x_i$ is computed, update $e(S)=e(S)+x_i$. Remove all the edges incident to $v_i$ from $G_i$. Denote the new graph by $G_{i+1}=(V_{i+1},E_{i+1})$.
\end{enumerate}

\par Intuitively, the above procedure first ``counts'' the number of edges in $G_i$ that belongs to $\bar{E}$ and is connected to the target vertex $v_i$, namely, the $|\{(v_i,j)\in E_i \cap \bar{E}:j\in N\}|$ term in \eqref{eq:inter9}. Then, if the target vertex $v_i\in S_{-c}$, we compute one more term in the following way. For each edge $(u,v_i)\in E_i\cap \tilde{E}$, if the following inequality holds, we also count this specific $(u,v_i)$ edge:

\begin{align}
|\{(u,j)\in E_i:j\in N\}|>|N_{-c}|.
\end{align}
That is, we check how many edges (including those in $\bar{E}\cap E_i$ and in $\tilde{E}\cap E_i$) are still connected to $u$. We count the single edge $(u,v_i)$ if there are still at least $(|N_{-c}|+1)$ edges in $E_i$ that are connected to $u$. Collectively, this additional counting mechanism for the case of $v_i\in S_{-c}$ gives the second term in \eqref{eq:inter9}. After counting the edges incident to $v_i$, we remove those edges from  $E_i$ so that in the future counting rounds (rounds $>i$) we do not double count the edges in any way.

\begin{claim}\label{clm:proc_count}
After finishing the subroutine {\sc Count}, the final $e(S)$ value is exactly the value of $\mathsf{ a.count}$.
\end{claim}
\begin{IEEEproof} The proof of the above claim is as follows. We first note that in the subroutine, we order the nodes in $S$ in the specific order such that all nodes in $S_{-c}$ are placed last. Therefore, in the beginning of the subroutine {\sc Count}, all the $v_i$ vertices do not belong to $S_{-c}$. For that reason, the second term in \eqref{eq:inter9} is zero. Since $v_i\notin S_{-c}$, all the edges connected to $v_i$ are in $\bar{E}$. The first term of \eqref{eq:inter9} thus ensures that we count all those edges in this subroutine. Since we remove those counted edges in each step (from $G_i$ to $G_{i+1}$), we do not double count any of the edges. Therefore, before we start to encounter a vertex $v_i\in S_{-c}$, the subroutine correctly counts the number of edges incident to the $v_{j}$ for all $1\leq j<i$.

\par We now consider the second half of the subroutine, i.e., when $v_i\in S_{-c}$. We then notice that the subroutine still counts all those edges in $\bar{E}$ through the first term in \eqref{eq:inter9}. The only difference between {\sc Count} and a regular counting procedure is the second term in \eqref{eq:inter9}. That is, when counting any edge in $\tilde{E}$, we need to first check whether the total number of edges in $G_i$ incident to $u$ is greater than $|N_{-c}|$. To explain why we have this {\em conditional counting} mechanism,
we notice that in the original graph $G$, each node $u\in N_0$ has $|\{(u,j)\in \bar{E}:j\in N\}|=d$ and $|\{(u,j)\in\tilde{E}:j\in N\}|=|N_{-c}|$. Therefore, the total number of edges connected to $u$ is $|\{(u,j)\in E:j\in N\}|=d+|N_{-c}|$. Note that during the counting process, those counted edges are removed from the graph during each step. Since $G_i$ is the remaining graph after removing all those counted edges in the previous $(i-1)$ steps, if we still have $|\{(u,j)\in E_i:j\in N\}|>|N_{-c}|$, then it means that we have only removed strictly less than $(d+|N_{-c}|)-|N_{-c}|=d$ number of edges in the previous $(i-1)$ counting rounds. The above argument thus implies that in the previous $(i-1)$ counting rounds, we have only counted $<d$ edges that are incident to node $u$.

\par Without loss of generality, we assume that $u$ is the $m$-th node of $N_0$. Then it means that the $a_m$ value (the number of edges connected to $u$) computed thus far (until the beginning of the $i$-th counting round) is still strictly less than $d$. Therefore, when computing the objective value $\mathsf{a.count}=a_0+\sum_m\min(a_m,d)$, the to-be-considered edge $(v_i,u) $ in the second term of \eqref{eq:inter9} will increment $a_m$ value by 1 and thus increment $\mathsf{a.count}$ by 1. Since our goal is to correctly compute the $\mathsf{a.count}$ value by this subroutine, the subroutine needs to include this edge into the computation, which leads to the second term in \eqref{eq:inter9}.

\par On the other hand, if the total number of edges in $G_i$ that are adjacent to $u$ is $\leq |N_{-c}|$, it means that we have removed $\geq (d+|N_{-c}|)-|N_{-c}|=d$ number of edges in the previous counting rounds. That is, when counting those edges adjacent to $u$, we have already included/encountered $\geq d$ such edges in the previous $(i-1)$ rounds. As a result, the corresponding $a_m$ value is $\geq d$. Therefore, when computing the objective value $\mathsf{a.count}=a_0+\sum_m\min(a_m,d)$, the to-be-considered edge $(v_i,u) $ will increment the value of $a_m$ by 1 but {\em will not} increment the $\mathsf{a.count}$ value. In the subroutine {\sc Count}, we thus do not count the edges in $\tilde{E}_i$ anymore, which leads to the second term in \eqref{eq:inter9}.

\par The new constraint put in Step~3 thus ensures that the final output $e(S)$ is the value of $\mathsf{a.count}$.
\end{IEEEproof}

\par We now need to prove that for any set $S$ of $k$ nodes, the corresponding $e(S)\geq\mathcal{M}$. Assuming this is true, we can then invoke Property~2, which guarantees that we can reconstruct the $\mathcal{M}$ packets of the original file from the coded packets stored in $S$.


\par To prove that $e(S)\geq\mathcal{M}$, we need the following claim.

\begin{claim} \label{clm:FFR}
For any arbitrarily given set $S$, there exists an $\tilde{\mathbf{r}} \stackrel{\Delta}{=}(\tilde{r}_1,\cdots, \tilde{r}_k)\in \{1,2,\cdots, n\}^k$ such that
\begin{align}\label{eq:claim2}
e(S)=\sum_{i=1}^k(d-z_i(\tilde{\mathbf{r}})),
\end{align}
where $z_i(\cdot)$ is a function $z_i:\{1,\cdots, n\}^k\mapsto \mathbb{N}$ defined as $z_i(\mathbf{r})=|\{a\in D_{r_i}:\exists j<i, a=r_j\}|$,  where $\mathbb{N}$ is the set of all positive integers and $D_{r_i}$ is the helper set of node $r_i$ in our FFR code construction. Additional explanation of the function $z_i(\cdot)$ can be found in \cite[Lemma~8]{ahmad2018can}. 
\end{claim}

Using the above claim, we have
\begin{align}
\mathsf{a.count}&=e(S)=\sum_{i=1}^k(d-z_i(\tilde{\mathbf{r}}))\label{eq:FFR_1}\\
&\geq \min_{\mathbf{r}\in \{1,\cdots, n\}^k} \sum_{i=1}^k(d-z_i(\mathbf{r}))\label{eq:FFR_2}\\
&=\sum _{i=1}^{k}(d-y_i)\label{eq:FFR_3}\\
&= \mathcal{M}\label{eq:FFR_4}.
\end{align}
where \eqref{eq:FFR_1} follows from Claim~\ref{clm:FFR}; \eqref{eq:FFR_2} follows from taking the minimum operation; \eqref{eq:FFR_3} follows from \cite[Proposition~6]{ahmad2018can} and Lemmas~8 to 11 of the proof of \cite[Proposition~5]{ahmad2018can}; and \eqref{eq:FFR_4} follows from \eqref{eq:file_size_construction}. By Property~2, we have thus proved that the $k\cdot d$ packets stored in any set of $k$ nodes can be used to jointly reconstruct the original file of size $\mathcal{M}$.

\par The proof of Claim~\ref{clm:FFR} is provided in Appendix~\ref{app:clm3}. The proof that the FFR codes can protect against $(n-k)$ simultaneous failures is hence complete.




\end{IEEEproof}

\section{The Extension of The FFR Codes}\label{sec:FFR_extension}
In our FFR codes, the helper selection scheme is based on the FHS scheme proposed in \cite{ahmad2018can}. In \cite{ahmad2018can}, the FHS scheme is also extended to a new scheme called, family-plus scheme. In this section, we introduce some basic notation/concepts of the family-plus helper selection scheme and then discuss how we can generalize the FFR codes in this work so that we can also replace the FHS scheme in the FFR codes by the new family-plus scheme.
\subsection{The Family-Plus Helper Selection Scheme}
\par The family-plus helper selection scheme is an extension of the FHS scheme for $n\gg d$. In family-plus helper selection, the $n$ nodes are grouped into several disjoint groups of $2d$ nodes and one disjoint group of $n_{\text{remain}}$ nodes. The first type of groups is termed the regular group while the second type is termed the remaining group. If there has to be one remaining group (when $n\bmod (2d)\neq 0$), then it is enforced that the size of the remaining group is as small as possible but still satisfying $n_\text{remain}\geq 2d+1$. After the partitioning, the FHS scheme is applied to the individual groups. Specifically, if a newcomer belongs to the first group, then all its helpers are chosen within the same group according the rules of the original FHS scheme. Since each regular group is of size $2d$ nodes and each remaining group must satisfy $n_\text{remain}\geq 2d+1$, one can easily verify that whenever $n\leq 4d$, then there is no regular group and only 1 remaining group. As a result, the family-plus scheme collapses back to the original FHS scheme. On the other hand, when $n\geq 4d+1$, then there will be multiple groups and the family-plus scheme differs from the FHS scheme. 

\par The file size that can be protected at the MBR point of the family-plus helper selection scheme was found in \cite{ahmad2018can} to be
\begin{align}\label{eq:mbr_plus}
\mathcal{M}=\Bigg(&1_{\{n\bmod(2d)\neq 0\}}\cdot \sum_{i=0}^{\min(k,2d-1)-1}\left(d-i+\left\lfloor\frac{i}{2}\right\rfloor\right)+\nonumber\\
&d^2\left\lfloor \frac{(k-n_l)^+}{2d}\right\rfloor+ \sum_{i=0}^{q}\left(d-i+\left\lfloor\frac{i}{2}\right\rfloor\right)\Bigg) \beta,
\end{align}
where
\begin{align}
&q=((k-n_l)^+\bmod(2d))-1, \text{ and}\nonumber\\ 
&n_l=
\begin{cases}
n_{\text{remain}},& \text{ if } n\bmod(2d)\neq 0\\
0,& \text{ otherwise}.
\end{cases}\nonumber
\end{align}
In \cite{ahmad2018can}, it was proved that for any $(n,k,d)$ values, \eqref{eq:mbr_plus} is always no less than \eqref{eq:mbr}. That is, the family-plus scheme improves upon the FHS scheme regardless whether $n\leq 4d-1$ or $n\geq 4d$.

\subsection{The FFR Codes Based on the Family-Plus Scheme}
\par The FFR codes described above can be modified and used to construct an explicit exact-repair code that can achieve the MBR point of the family-plus helper selection scheme. This is achieved by first applying the same inner code graph construction of the above FFR codes to each group of the family-plus helper selection scheme, i.e., the edge representation of each group consists of the two edge sets $\bar{E}$ and $\tilde{E}$. Then, since the repair of the family-plus scheme occurs within each group separately, for the outer code, we enforce Property~1 for each individual group so that we can maintain the exact-repair property. Finally, we need to ensure that any subset of $k$ nodes (which could be across multiple groups) can be used to reconstruct the original file. Therefore, we have to ensure that the outer code satisfies a modified version of Property~2. 
\par In the following we briefly describe how to do this modification with a slight abuse of notation. Recall that in the family-plus helper selection scheme, only the remaining group could possibly have an incomplete family. Denote the set of incomplete family nodes in the remaining group by $N_0$ and the graph of the remaining group by $G_{\text{remain}}=(V_{\text{remain}},E_{\text{remain}})$. The new property imposed on the packets becomes
{\bf Modified Property~2:} Index the vertices in $N_0\subset V_{\text{remain}}$ by $\{u_1,u_2,\cdots,u_{|N_0|}\}$. For any given subset of the total packets (across all groups) and any given $m$ satisfying $1\leq m\leq |N_0|$, define $a_m$ as the number of packets in this subset that correspond to the edges in $E_{\text{remain}}=\bar{E}_{\text{remain}}\cup\tilde{E}_{\text{remain}}$ that are incident to vertex $u_m\in N_0$. Define $a_0$ as the number of the other packets in this subset, i.e., those packets not corresponding to any edges that are incident to $N_0$. Define $\mathsf{a.count}\stackrel{\Delta}{=} a_0+\sum_{m=1}^{|N_0|}\min(a_m,d)$. The modified Property~2 enforces that we must be able to reconstruct the original file of size $\mathcal{M}$ if $\mathsf{a.count}\geq \mathcal{M}$. 

\par We can again use the concept of random linear network coding to prove the existence of a code satisfying Property~1 and the Modified Property~2 in a similar way as in Lemma~\ref{lem:FFR_existence}. The correctness of the proposed FFR codes for family-plus helper selection schemes can be proved in a similar way as when proving the correctness for FHS schemes provided in Section~\ref{sec:FFR_proof}. We omit the detailed proofs since they are simple extensions of the proofs provided for the FHS scheme with only the added notational complexity of handling different groups of nodes in the family-plus helper selection schemes. 

\section{conclusion} \label{sec:conc}
\par In this paper, we have presented a new class of codes that we term \emph{flexible fractional repetition (FFR) codes}. These codes possess several important properties: (i) they achieve the MBR point of the FHS scheme and close the loop of the graph-based necessary and sufficient  condition of the benefits of helper selection derived in \cite{ahmad2018can}; (ii) the proposed FFR codes are exact-repair codes and for the most part admit the repair-by-transfer property; and (iii) their construction utilizes a new code-construction technique that generalizes the existing FR codes for arbitrary network parameters. One future direction is to further generalize the proposed FFR codes for the multiple failures scenario in a way similar to the existing results in \cite{el2010fractional,olmez2013replication}.

\appendices

\section{Proof of Lemma~\ref{lem:FFR_existence}} \label{app:FFR_existence}

\begin{figure}[h!]
\centering
\includegraphics[width=0.45\textwidth]{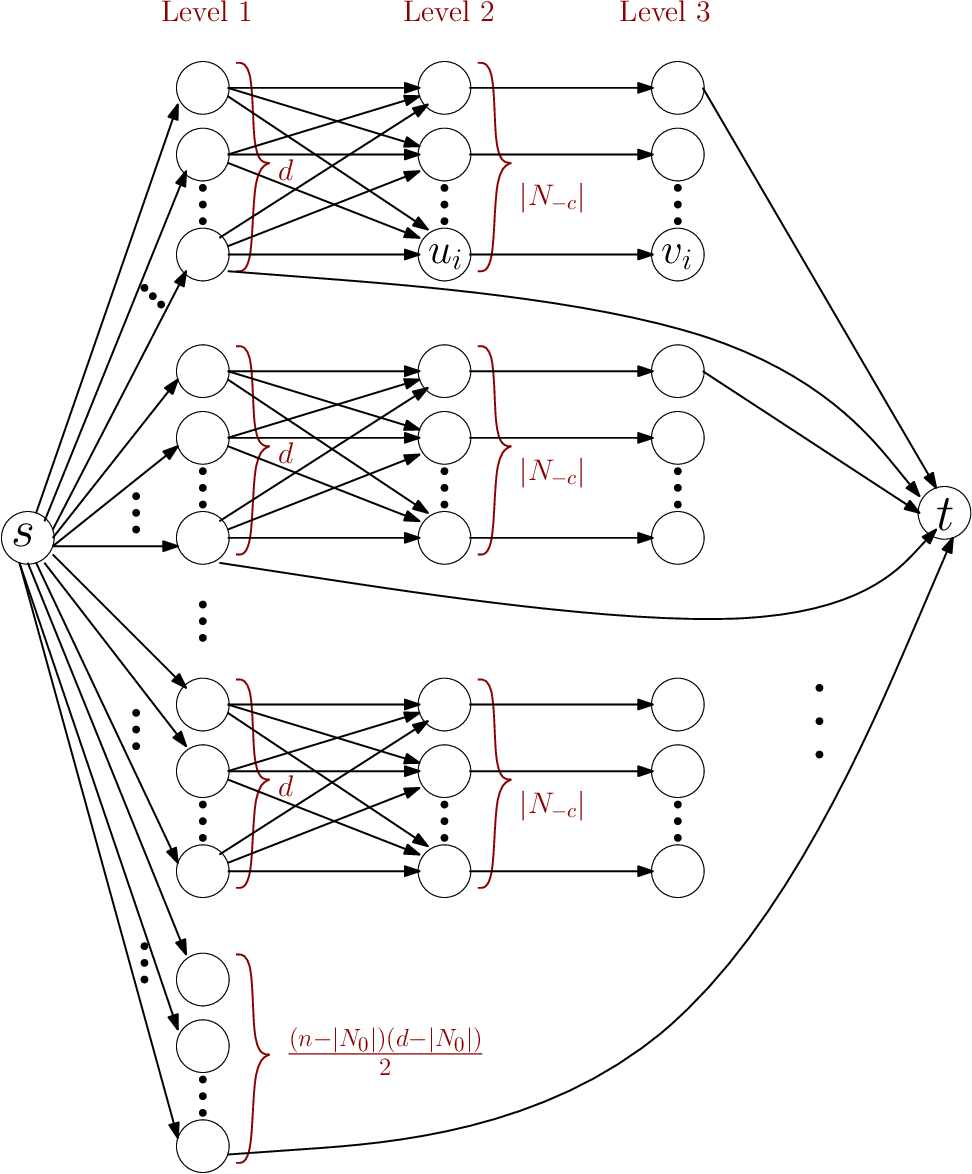}
\caption{The graph of the proof of Lemma~\ref{lem:FFR_existence}.}
\label{fig:FFR_existence}
\end{figure}

To prove this lemma, we model the problem using a finite directed acyclic graph and then we invoke the results from random linear network coding \cite{ho2006random}. The graph has a single source vertex $s$ that is incident to $|\bar{E}|=|{\sf IJ}^{[1]}|+|{\sf IJ}^{[2]}|=\frac{(n-|N_0|)(d-|N_0|)}{2}+d|N_0|$ other vertices with edges of capacity 1. We call these vertices \emph{level 1} vertices. Among these level 1 vertices, we focus on a subset of $d|N_0|$ vertices and partition it into $|N_0|$ disjoint groups and each group consists of $d$ arbitrarily chosen distinct vertices. The intuition is that each group of them is associated with a vertex in $N_0$. See Fig.~\ref{fig:FFR_existence} for illustration.

\par Now, in addition to the source $s$ and the level 1 vertices, we add $|N_0|\cdot |N_{-c}|$ new node pairs $(u_i,v_i)$ for all $1\leq i\leq |N_0|\cdot |N_{-c}|$. Each $(u_i,v_i)$ is connected by an edge of capacity 1. We call the $u_i$ nodes, level 2 vertices and the $v_i$ nodes level 3 vertices. We partition the new node pairs (edges) into $|N_0|$ groups and each group consists of $|N_{-c}|$ edges. We then associate each group of $|N_{-c}|$ edges to one group of $d$ level 1 vertices created previously. See Fig.~\ref{fig:FFR_existence} for illustration. Finally, for the level 1, level 2, and level 3 vertices belonging to the same group (there are $|N_0|$ groups in total), we connect all the level 1 vertices in this group and all the level 2 vertices in this group by an edge with infinite capacity.

\par We now describe the relationship of the newly constructed graph in Fig.~\ref{fig:FFR_existence} to the graph representation of the FFR code. For easier reference, we use the graph in Fig.~\ref{fig:FFR_existence} to refer to the newly constructed graph; and use the graph in Fig.~\ref{fig:ex3} to refer to the graph representation of the FFR codes. There are $|N_0|$ groups in the graph of Fig.~\ref{fig:FFR_existence} and each group corresponds to one node in $N_0$ of the graph of Fig.~\ref{fig:ex3}. We notice that there are $|\bar{E}|=\frac{(n-|N_0|)(d-|N_0|)}{2}+d|N_0|$ number of level 1 vertices in the graph of Fig.~\ref{fig:FFR_existence} and $|\bar{E}|=\frac{(n-|N_0|)(d-|N_0|)}{2}+d|N_0|$ number of edges in $\bar{E}$ of the graph of Fig.~\ref{fig:ex3}. As a result, we map each level 1 vertex bijectively to an edge in $\bar{E}$ in a way that each group of the level-1 vertices in Fig.~\ref{fig:FFR_existence} (totally $|N_0|$ groups) corresponds to the $d$ edges in $\bar{E}$ that are connected to the same node in $N_0$ of Fig.~\ref{fig:ex3}.  

\par There are $|N_0|\cdot |N_{-c}|$ number of level 3 vertices in the graph of Fig.~\ref{fig:FFR_existence} and there are $|N_0|\cdot |N_{-c}|$ number of $\tilde{E}$ edges in the graph of Fig.~\ref{fig:ex3}. As a result, we map each level 3 vertex bijectively to an edge in $\tilde{E}$ in a way that each group of the level-3 vertices in Fig.~\ref{fig:FFR_existence} (totally $|N_0|$ groups) corresponds to the $|N_{-c}|$ edges in $\tilde{E}$ that are connected to the corresponding node in $N_0$ of Fig.~\ref{fig:ex3}.

\par We now focus on how to encode over the graph of Fig.~\ref{fig:FFR_existence} and then use the above mapping to convert it to a coding scheme over the graph of Fig.~\ref{fig:ex3}. Assume that source $s$ has a file of $\mathcal{M}$ packets. We perform random linear network coding (RLNC) \cite{ho2006random} on the graph of Fig.~\ref{fig:FFR_existence} assuming a sufficiently large finite field $\text{GF}(q)$ is used. Specifically, for any level-1 vertex $u$, the corresponding $(s,u)$ is a random mixture of all $\mathcal{M}$ packets. For each $(u_i, v_i)$ edge connecting a level-2 vertex $u_i$ and a level-3 vertex $v_i$, it carries a linear combination of all coded $(s,u)$ edges that are incident\footnote{Technically, we should say all $(s,u)$ edges that are {\em upstream} of $(u_i,v_i)$. However, since the edges connecting level-1 and level-2 vertices are of infinite capacity, we use the word {\em incident} in a loose sense.} to $(u_i,v_i)$. After the encoding over Fig.~\ref{fig:FFR_existence} is fixed, we can immediately construct the corresponding encoding scheme over Fig.~\ref{fig:ex3} based on the aforementioned mapping. For example, for a level-2 vertex $u$ and a level-3 vertex $v$, if edge $(u,v)$ belongs to the $i_0$-th group in Fig.~\ref{fig:FFR_existence} and $v$ is the $j_0$-th level 3 vertex in this group, then, we assign the coded packets on the edge $(u,v)$ to the edge $e\in \tilde{E}$ (in the graph of Fig.~\ref{fig:ex3}) that connects the $i_0$-th node in $N_0$ and the $j_0$-th node in $N_{-c}$.

\par In the following, we will prove that with a sufficiently large $\GF(q)$ the above code construction (from the RLNC-based code in the graph of Fig.~\ref{fig:FFR_existence} to the FFR codes in the graph of Fig.~\ref{fig:ex3}) satisfies Properties~1 and 2 with close-to-one probability. 

\par We first prove that our construction satisfies Property~1 with probability one. To that end, we notice that any coded packet $\tilde{P}_{(i_0,j_0)}$ corresponding to some $(i_0,j_0)\in {\sf IJ}^{[3]}$ in the graph of Fig.~\ref{fig:ex3} is now mapped from a $(u,v)$ edge in Fig.~\ref{fig:FFR_existence} where $u$ is a level 2 vertex; $v$ is a level 3 vertex; $(u,v)$ belongs to the $i_0$-th group in Fig.~\ref{fig:FFR_existence}; and $v$ is the $j_0$-th level 3 vertex in this group. By the graph construction in Fig.~\ref{fig:FFR_existence}, such a coded packet is a linear combination of the coded packets in Fig.~\ref{fig:FFR_existence} from source $s$ to vertex $\tilde{u}$ where the $\tilde{u}$ vertices are the level-1 vertices corresponding to the $i_0$-th group. Since those packets along $(s,\tilde{u})$ are the $P_{(j_1,i_0)}$ packets for all $j_1$ satisfying $(j_1,i_0)\in {\sf IJ}^{[2]}$ in the graph of Fig.~\ref{fig:ex3}, we have thus proved Property~1: Specifically, any coded packet $\tilde{P}_{(i_0,j_0)}$ corresponding to some $(i_0,j_0)\in {\sf IJ}^{[3]}$ is a linear combination of the packets $P_{(j_1,i_0)}$ for all $j_1$ satisfying $(j_1,i_0)\in {\sf IJ}^{[2]}$.

\par To prove that the above construction satisfies Property~2 with close-to-one probability, for any edge set subset of edges in the graph of Fig.~\ref{fig:ex3}  with the corresponding $\mathsf{a.count}$ value satisfying $\mathsf{a.count}\geq \mathcal{M}$, we place a sink node $t$ in the graph of Fig.~\ref{fig:FFR_existence} that connects to the corresponding set of level 1/level 3 vertices in Fig.~\ref{fig:FFR_existence} using edges of infinite capacity. See Fig.~\ref{fig:FFR_existence} for illustration of one such $t$. One can quickly verify that the min-cut-value from the source $s$ to the sink $t$ in the graph of Fig.~\ref{fig:FFR_existence} is indeed the $\mathsf{a.count}$ value computed from the given subset of edges in the graph of Fig.~\ref{fig:ex3}. As a result, with a sufficiently large finite field $\text{GF}(q)$, sink $t$ in Fig.~\ref{fig:FFR_existence} can successfully reconstruct the original file with close-to-one probability. Since the sink $t$ accesses only level 1 and level 3 vertices, the $P_{(i,j)}$ packets in the graph of Fig.~\ref{fig:ex3} that correspond to the level 1 vertices in the graph of Fig.~\ref{fig:FFR_existence} and the $\tilde{P}_{(i,j)}$ packets in the graph of Fig.~\ref{fig:ex3} that correspond to the level 3 vertices in the graph of Fig.~\ref{fig:FFR_existence} jointly can reconstruct the original file of size $\mathcal{M}$. Property~2 is thus also satisfied. Since, there are at most ${|E|\choose \mathcal{M}}$ different ways of choosing the sink $t$,\footnote{Since ultimately we are only interested in reconstructing the file from any $k$ nodes, we actually only need to consider ${n\choose k}$ ways of choosing the sink $t$, which can further improve the probability lower bound.} a very loose outer bound of the success probability is 
\begin{align}
\prob(\text{The RLNC construction satisfies Lemma~\ref{lem:FFR_existence}})\geq 1-\frac{{|E|\choose \mathcal{M}}}{q}.
\end{align} 
\par By the above arguments, the proof of Lemma~\ref{lem:FFR_existence} is complete.

\section{Proof of Claim~\ref{clm:FFR}} \label{app:clm3}
In order to prove Claim~\ref{clm:FFR}, we will need the following fact.
\begin{claim}\label{clm:edges_count}
Suppose there exists a node $a\in S_{-c}$ and a node $b\in N_c\backslash S_c$. Define a new set of nodes $S'\stackrel{\Delta}{=}(S\cup\{b\})\backslash a$. That is, we remove node $a$ from $S$ but add a new node $b$ in $S$ that satisfies $b\in N_c$. Then
\begin{align} \label{eq:claim1}
e(S)=e(S').
\end{align}
That is, running the subroutine {\sc Count} on both $S$ and $S'$ will lead to the same final output value. 
\end{claim}

\begin{IEEEproof} We consider {\sc Count} for the set $S'$ and we denote nodes in $S'$ that belong to $N_i$ by $S_i'\stackrel{\Delta}{=}S'\cap N_i$. To avoid confusion when $S'$ is used as input to the subroutine {\sc Count}, we call the new graphs during the counting steps of {\sc Count} by $G'_{i}=(V'_i,E'_i)$, the new vertices by $v'_{i}$, and the new $x_i$ by $x'_i$. Since the subroutine {\sc Count} can be based on any sorting order of nodes in $S$ (and in $S'$) as long as those nodes in $N_{-c}$ come last, we assume that the nodes in $S$ are sorted in a way that node $a$ is the very first node in $S_{-c}$. For convenience, we say that node $a$ is the $i_0$-th node in $S$ and we assume that all the first $(i_0-1)$-th nodes are not in $S_{-c}$ and all the nodes following the $(i_0-1)$-th node are in $S_{-c}$. That is, $i_0=|S|-|S_{-c}|+1=k+1-|S_{-c}|$.  We now use the same sorting order of $S$ and apply it to $S'$. Specifically, the $i$-th node of $S$ is the same as the $i$-th node in $S'$ except for the case of $i=i_0$. The $i_0$-th node of $S'$ is set to be node $b$. One can easily check that the sorting orders of $S$ and $S'$ both satisfy the required condition in Step~1 of the subroutine {\sc Count}.

\par We will run {\sc Count} on both $S$ and $(S\cup \{b\})\backslash a$ in parallel and compare the resulting $e(S)$ and $e((S\cup \{b\})\backslash a)$.

\par It is clear that in rounds 1 to $(i_0-1)$, the subroutine {\sc Count} behaves identically when applied to the two different sets $S$ and $S'=(S\cup \{b\})\backslash a$ since their first $(i_0-1)$ vertices are identical. We now consider the $i_0$-th round and argue that the total number of edges in $E'_{i_0}$ incident to $v'_{i_0}$ is equal to the total number of edges incident to $v_{i_0}$ in $E_{i_0}$. Recall that $b$ and $a$ have the same helper sets since they are from the same complete family. Specifically, the edges in $E$ incident to $v_{i_0}=a\in S_{-c}$ that have been counted in the first $(i_0-1)$ rounds are of the form $(u,a)$ for all $u\in\{v_1,v_2,\cdots,v_{i_0-1}\}\cap (S_0\cup S_1\cup \cdots \cup S_{c-1})$. Also note that in the original graph $G$, there are exactly $d$ edges incident to node $a\in S_{-c}$ (some of them are in $\bar{E}$ and some of them in $\tilde{E}$). Therefore, in $E_{i_0}$ (after removing those previously counted edges), there are $(d-|\{v_1,v_2,\cdots,v_{i_0-1}\}\cap (S_0\cup S_1\cup \cdots \cup S_{c-1})|)$ number of edges that are incident to $v_{i_0}$.

\par Similarly, the edges in $E_{i_0}'$ incident to $v_{i_0}'=b\in S_{c}'$ that have been counted previously are of the form $(u,b)$ for all $u\in \{v_1,v_2,\cdots,v_{i_0-1}\}\cap (S_0\cup S_1\cup \cdots \cup S_{c-1})$ since $v_i'=v_i$ for $1\leq i\leq i_0-1$ and $S_x'=S_x$ for $0\leq x\leq c-1$. Also note that, in the original graph $G'$, there are exactly $d$ edges incident to node $b\in S_{c}'$ (all of them are in $\bar{E'}$). Therefore, in $E_{i_0}'$ (after removing those previously counted edges), there are $(d-|\{v_1,v_2,\cdots,v_{i_0-1}\}\cap (S_0\cup S_1\cup \cdots \cup S_{c-1})|)$ number of edges that are incident to $v_{i_0}'=b$.

\par We now argue that all the edges in $E_{i_0}$ that are incident to $a$ will contribute to the computation of $x_{i_0}$. The reason is that node $a$ is the first vertex in $S_{-c}$. Therefore, when in the $i_0$-th counting round, no edge of the form $(u,v)$ where $u\in N_0 \backslash S_0$ and $v\in N_{-c}$ has ever been counted in the previous $(i_0-1)$ rounds. Also, since we choose $b\in N_c\backslash S$ to begin with, when running {\sc Count} on $S$, for all $u\in N_0\backslash S_0$ at least one edge, edge $(u,b)$, is not counted during the first $(i_0-1)$ rounds. As a result, for any $u\in N_0\backslash S_0$, in the $i_0$-th round, at least $|\{(u,v): v\in N_{-c}\}|+1=|N_{-c}|+1$ edges incident to $u$ are still in $E_{i_0}$ (not removed in the previous $(i_0-1)$ rounds). This thus implies that the second term of \eqref{eq:inter9} will be non-zero. Therefore, at the $i_0$-th iteration of Step~3 of {\sc Count}, all the edges in $E_{i_0}$ incident to $v_{i_0}=a$ are counted. The $x_{i_0}$ value computed in \eqref{eq:inter9} thus becomes $x_{i_0}= d-|\{v_1,v_2,\cdots,v_{i_0-1}\}\cap (S_0\cup S_1\cup \cdots \cup S_{c-1})|$.

\par The previous paragraph focuses on the $i_0$-th round when running the subroutine {\sc Count} on $S$. We now consider the $i_0$-th round when running {\sc Count} on $S'$. We argue that all the edges in $E'_{i_0}$ that are incident to $b$ will contribute to the computation of $x'_{i_0}$. The reason is that node $b\in S_{c}'$. Therefore, all edges incident to $b$ belong to $\bar{E'}$. As a result, all the edges in $E'_{i_0}$ that are incident to $b$ will contribute to the computation of $x'_{i_0}$ through the first term in \eqref{eq:inter9}. We thus have $x'_{i_0}= d-|\{v_1,v_2,\cdots,v_{i_0-1}\}\cap (S_0\cup S_1\cup \cdots \cup S_{c-1})|$.

Since $x_{i_0}=x'_{i_0}$, we thus have $e(S)=e(S')$ after the first $i_0$ counting rounds.

\par We now consider rounds $(i_0+1)$ to $k$. We observe that by our construction $v'_i=v_i\in S'_{-c}\subset S_{-c}$ for $i_0+1 \leq i\leq k$. Moreover, since $v_{i_0}=a\in S_{-c}$ and $v'_{i_0}=b\in S_c'$, both vertices $a$ and $b$ are initially not connected to any vertices in $S_{-c}$ and $S_{-c}'$ respectively (those $v_i$ and $v'_i$ with $i_0+1\leq i\leq k$) since vertices of the same family are not connected. Therefore, replacing the $i_0$-th node $v_{i_0}=a$ by $v_{i_0}'=b$ will not change the value of the first term in \eqref{eq:inter9} when computing $x_i$ for the $i$-th round where $i_0+1\leq i\leq k$.

\par We now consider the second term of \eqref{eq:inter9}. For any $u\in S_0$, any edge incident to $u$ has been counted in the first $(i_0-1)$ rounds since we assume that when we are running {\sc Count} on the $S$ set, we examine the nodes in $S_{-c}$ in the very last. Therefore, there is no edge of the form $(v_i,u)$ in $E_i$ (resp. $(v_i',u) \in E_i'$) with $u\in S_0$ since those edges have been removed previously. Therefore, the summation over $u\in N_0$ can be replaced by $u\in N_0\backslash S_0$ during the $i_0$-th round to the $k$-th round. On the other hand, for any $u\in N_0\backslash S_0$, if there is an edge connecting $(a,u)\in \tilde{E}$, then by our construction there is an edge $(b,u)\in \bar{E}$. Therefore, in the $i_0$-th round, the same number of edges incident to $u$ is removed regardless whether we are using $S$ as the input to the subroutine {\sc Count} or we are using $S'$ as the input to the subroutine {\sc Count}. As a result, in the beginning of the $(i_0+1)$-th round, for any $u\in N_0$, we have the following equality
\begin{align}
|\{(u,j)\in E_i:j\in N\}|=|\{(u,j)\in E_i':j\in N\}|\label{new:counting-eq}
\end{align}
when $i=i_0+1$. Moreover, for any $u\in N_0\backslash S_0$, we remove one and only one edge $(u,v_i)$ in the $i$-th round, regardless whether we are counting over $S$ or over $S'$. Since $v_i=v_i'$ for all $i=i_0+1$ to $k$, we have \eqref{new:counting-eq} for all $i=i_0+1$ to $k$ as well. The above arguments thus prove that the second term of \eqref{eq:inter9} does not change regardless whether we count over $S$ or $S'$. As a result, $x'_i=x_i$ for $i_0+1\leq i\leq k$. Since $e(S)=e(S')$ for all $k$ rounds of the counting process, we have thus proved \eqref{eq:claim1}.
\end{IEEEproof}

\par We now turn our attention back to proving Claim~\ref{clm:FFR}. For any node set $S$, by iteratively using Claim~\ref{clm:edges_count}, we can construct another node set $S'$ such that $e(S)=e(S')$ while either (Case~i) $S'_{-c}=\emptyset$; or (Case~ii) $S'_{-c}\neq \emptyset$ and $S'_c=N_c$. As a result, we can assume without loss of generality that we have either (Case~i) $S_{-c}=\emptyset$; or (Case~ii) $S_{-c}\neq \emptyset$ and $S_c=N_c$ to begin with.

\par We first consider the former case. Let $\tilde{\mathbf{r}}$ be any vector in $R$ such that its $\tilde{r}_i=v_i$ for $1\leq i\leq k$, i.e., $\tilde{r}_i$ equals the node index of the vertex $v_i$. We will run the subroutine {\sc Count} sequentially for $i=1$ to $k$ and compare the increment of $e(S)$ in each round, denoted by $x_i$ in \eqref{eq:inter9}, to the $i$-th term $(d-z_i(\tilde{\mathbf{r}}))$ in the summation of the right-hand side of \eqref{eq:claim2}. Consider the $i$-th round of counting for some $1\leq i\leq k$, and assume that the corresponding vertex $v_{i}$ belongs to the $y$-th family, i.e., $v_{i} \in N_y$. Since $S_{-c}=\emptyset$ in this case, we have $v_{i}\notin S_{-c}$ and the second term in \eqref{eq:inter9} is always 0. Therefore, the procedure {\sc Count} is indeed counting the number of edges in $\bar{E}$ that are incident to $S$ without the special conditional counting mechanism in the second term of \eqref{eq:inter9}. Therefore, we have
\begin{align} x_{i}&=|\{(v_{i},j)\in E_{i}\cap\bar{E}:j\in N\}|\nonumber\\
&=d-|\{v_j\notin N_{y}:v_j\in S, 1\leq j\leq i-1\}|,  \label{eq:xi_zi}
\end{align}
where $d$ is the number of $\bar{E}$ edges in the original graph $G$ that are incident to $v_{i}$ and $|\{v_j\notin N_{y}:v_j\in S, 1\leq j\leq i-1\}|$ is the number of edges removed during the first $(i-1)$ counting rounds. On the other hand, we have 
\begin{align}
v_j\in D_y\Leftrightarrow v_j\in D_y\backslash N_{-c}\Leftrightarrow v_j \in N\backslash (N_y\cup N_{-c})
\end{align}
where the first equality follows from that $S_{-c}=\emptyset$ implies $v_j\notin N_{-c}$; the second equality follows from the FHS construction that $D_y\backslash N_{-c}=N\backslash (N_{-c}\cup N_{y})$ for any family index $y\neq -c$.  By the definition of function $z_i(\cdot)$, our construction of $\tilde{\mathbf{r}}$ thus always has $|\{v_j\notin N_y:v_j\in S, 1\leq j\leq i-1\}|=z_{i}(\tilde{\mathbf{r}})$. As a result, $x_{i}=(d-z_{i}(\tilde{\mathbf{r}}))$ for $i=1$ to $k$ and our explicitly constructed vector $\tilde{\mathbf{r}}$ satisfies \eqref{eq:claim2}.

\par We now turn our attention to the second case when $S_{-c}\neq \emptyset$ and $S_c=N_c$. Recall that  there are $k$ nodes in the set $S$. Let $\mathbf{r}$ be any vector in $R$ such that its $r_i=v_i$ for $1\leq i\leq k$. Define $j^*$ as the value that simultaneously satisfies (i) $k-|S_{-c }|\leq j^* \leq k$ and (ii) there are exactly $d$ entries in the first $j^*$ coordinates of $\mathbf{r}$ that are in $N\backslash N_0$. If no value satisfies the above two conditions simultaneously, set $j^*=k+1$. We now construct another vector $\tilde{\mathbf{r}}$ from $\mathbf{r}$ as follows: Replace the values of the $(j^*+1)$-th coordinate to the $k$-th coordinate of $\mathbf{r}$ by $n$, the node index of the last node in $N_0$ and denote the final vector by $\tilde{\mathbf{r}}$.

\par We will now prove that the above explicit construction of $\tilde{\mathbf{r}}$ satisfies the desired property in \eqref{eq:claim2}. The proof is divided into two cases:

\underline{Case~1:} There exists such a $j^*$ satisfying (i) and (ii). We will run the subroutine {\sc Count} again and compare $x_i$ to the $i$-th term $(d-z_i(\tilde{\mathbf{r}}))$.

We then observe the following facts:

\begin{enumerate}
\item In {\sc Count}, from $i=1$ to $(k-|S_{-c }|)$. For any $i$ in this range, we must have $FI(v_i)\neq -c$, i.e., the family index of node $v_i$ is not $-c$, since we run the subroutine {\sc Count} using a specific ordering of the nodes in $S$, which examines the nodes in $S_{-c}$ in the very last. As a result, the second term of \eqref{eq:inter9} is always zero. Therefore \eqref{eq:xi_zi} still holds. By the definition of function $z_i(\cdot)$, our construction of $\tilde{\mathbf{r}}$, and the fact that $1\leq i\leq k-|S_{-c}|$ (implying no $v_j\in S_{-c}$ for all $1\leq j\leq i-1$), we get $x_i=d-z_i(\tilde{\mathbf{r}})$ for all $1\leq i\leq k-|S_{-c}|$.

\item We now consider the case of $i=k-|S_{-c}|+1$ to $j^*$ of Step~3. For any $i$ in this range, we have $v_i\in S_{-c}$. We now argue that $|\{(u,j)\in E_i:j\in N\}|>|N_{-c}|$ for all edges $(u,v_i)\in E_i\cap \tilde{E}$ satisfying $u\in N_0$. The reason is that $(u,v_i)\in E_i$ implies that node $u$ is not counted in the previous $(i-1)$ rounds, i.e., $u\neq v_{i'}$ for all $1\leq i'\leq i-1$. Therefore, an edge of $(u,v)$ is removed if and only if there is a $v=v_j$ for some $v_j$ that is not in $N_0$. Since there are exactly $d$ vertices in $\{v_1,v_2,\dots,v_{j^*}\}$ that are not in $N_0$, it means that the first $(i-1)$ counting rounds where $1\leq i\leq j^*$ can remove at most $(d-1)$ edges incident to such a node $u$. Since node $u$ has $(d+|N_{-c}|)$ number of incident edges in the original graph $G$, we know that the inequality $|\{(u,j)\in E_i:j\in N \}|>|N_{-c}|$ must hold in the $i$-th round. As a result, the second term of \eqref{eq:inter9} is non-zero when $i=k-|S_{-c}|+1$ to $j^*$ and we can thus rewrite
\begin{align}
x_{i}&=|\{(v_{i},j)\in E_i:j\in N\}|\nonumber\\
&=d-|\{v_j\notin N_{c}\cup N_{-c}:v_j\in S, 1\leq j\leq i-1\}|.\nonumber
\end{align}
By the definition of function $z_i(\cdot)$ and our construction of $\tilde{\mathbf{r}}$, we get $x_i=d-z_i(\tilde{\mathbf{r}})$ for all $k-|S_{-c}|+1\leq i\leq j^*$.
\item We now consider the $(j^*+1)$-th to the $k$-th round of Step~3. We claim that
\begin{align}
x_i=d-|S_1\cup S_2\cup \cdots \cup S_c|
\end{align}
for those $j^*+1\leq i \leq k$. The reason behind this is the following. Since $j^*+1\leq i \leq k$, we have $v_i\in S_{-c}$. For any $u\in N_0\backslash S_0$ (those $u\in S_0$ have been considered in the first $(k-|S_{-c}|)$ rounds), there are $(d+|N_{-c}|)$ number edges incident to $u$ in the original graph $G$. On the other hand, since $i\geq j^*+1$ and by our construction, there are $d$ entries in the first $j^*$ coordinates of $\tilde{\mathbf{r}}$ that are are not in $N_0$, we must have removed at least $d$ edges incident to $u$ during the first $(i-1)$ counting rounds as discussed in the previous paragraph. Therefore, the number of incident edges in $E_i$ that are incident to $u\in N_0\backslash S_0$ must be $\leq |N_{-c}|$. The second term of \eqref{eq:inter9} is thus zero. As a result, the $x_i$ computed for $v_i$ will only include those edges in $E_i\cap \bar{E}$ incident to it. Since any $v_i\in S_{-c}$ only has $(d-|N_0|)$ number of edges in $\bar{E}$ to begin with, we have that
\begin{align}
x_i=(d-|N_0|)-|S_1\cup S_2 \cup \cdots \cup S_{c-1}|\nonumber
\end{align}
where $|S_1\cup S_2 \cup \cdots \cup S_{c-1}|$ is the number of edges in $\bar{E}$ that have been removed during the first $(i-1)$ rounds. Since $S_c=N_c$ in the scenario we are considering and since $|N_c|=|N_0|=n\bmod (n-d)$ in the FHS scheme, we can consequently rewrite $x_i$ as
\begin{align}
x_i=d-|S_1\cup S_2\cup \cdots \cup S_c|\nonumber
\end{align}
for $(j^*+1)\leq i\leq k$. Recall that in the newly constructed $\tilde{\mathbf{r}}$, the values of the $(j^*+1)$-th coordinate to the $k$-th coordinate are $n$, which belongs to $N_0$. Thus, by the definition of function $z_i(\cdot)$, we can see that each of these coordinates only contributes
\begin{align}
z_i(\tilde{\mathbf{r}})&=|\{\tilde{r}_j\in N\backslash (N_{-c}\cup N_0):1\leq j\leq i-1\}|\nonumber\\
&=|\{\tilde{r}_j\in N\backslash (N_{-c}\cup N_0):1\leq j\leq j^*\}|\label{eq:count-new-CCW1}\\
&=|S_1\cup S_2\cup \cdots \cup S_c|\nonumber
\end{align}
where \eqref{eq:count-new-CCW1} follows from the fact that in the construction of $\tilde{\mathbf{r}}$, the $(j^*+1)$-th to the $k$-th coordinates of $\tilde{\mathbf{r}}$ are always of value $n\in N_0$. Hence, we get $x_i=d-z_i(\tilde{\mathbf{r}})$ for $(j^*+1)\leq i\leq k$.
\end{enumerate}

We have proved for this case that $x_i=d-z_i(\tilde{\mathbf{r}})$ for $i=1$ to $k$. Therefore, we get \eqref{eq:claim2}.

\underline{Case~2:} No such $j^*$ exists. This means that one of the following two sub-cases is true. Case~2.1: even when choosing the largest $j^*=k$, we have strictly less than $d$ entries that are not in $N_0$. Case~2.2:  Even when choosing the smallest $j^*=k-|S_{-c}|$, we have strictly more than $d$ entries that are not in $N_0$.

\par Case~2.1 can be proved by the same arguments used in the previous proof of Case~1 (when proving the scenario of $k-|S_{-c}|+1\leq i\leq j^*$), which implies that we have $x_i=d-z_i(\tilde{\mathbf{r}})$ for all $1\leq i\leq k$. The proof of this case is complete.
\par Case~2.2 is actually an impossible case. The reason is that for any $1\leq i\leq k-|S_{-c}|$, there are exactly $|S_1|+|S_2|+\cdots + |S_c|$ nodes $v_i$ that are not in $N_0$, and we also have
\begin{align}
\sum_{m=1}^c |S_m|\leq\sum_{m=1}^c |N_m|=d,\nonumber
\end{align}
where the equality follows from our FHS construction. This, together with the observation that the first $(k-|S_{-c}|)$ coordinates of $\mathbf{r}$ are transcribed from the distinct nodes in $S_1\cup S_2\cup\cdots\cup S_c$, implies that we cannot have strictly more than $d$ entries that are not in $N_0$ in the first $(k-|S_{-c}|)$ coordinates of $\mathbf{r}$. Case 2.2 is thus an impossible case.
\par By the above arguments, the proof of Claim~\ref{clm:FFR} is complete.
\balance
\bibliography{paper}
\bibliographystyle{IEEEtranS}
\end{document}